\documentclass[aps,prl,preprint,superscriptaddress,showpacs,preprintnumbers,amsmath,amssymb]{revtex4}
\usepackage{graphicx}

\usepackage[usenames]{color}
\graphicspath{{ps/}}

\begin{document}

\def \BDDst  {\ensuremath{B^0\to D^{*\pm}D^\mp}}
\def \DP     {\ensuremath{D^+}}
\def \DM     {\ensuremath{D^-}}
\def \DstP   {\ensuremath{D^{*+}}}
\def \DstM   {\ensuremath{D^{*-}}}
\def \pis    {\ensuremath{\pi_{\rm slow}^+}}
\def \ltag   {\ensuremath{\ell_{\rm tag}}}
\def \ctheta {\ensuremath{\cos\theta}}
\def \calpha {\ensuremath{\cos\alpha}}
\def \Mbc    {\ensuremath{M_{\rm bc}}}
\def \DE     {\ensuremath{\Delta E}}
\def \Dt     {\ensuremath{\Delta t}}

\vspace*{-3\baselineskip}
\resizebox{!}{2.5cm}{\includegraphics{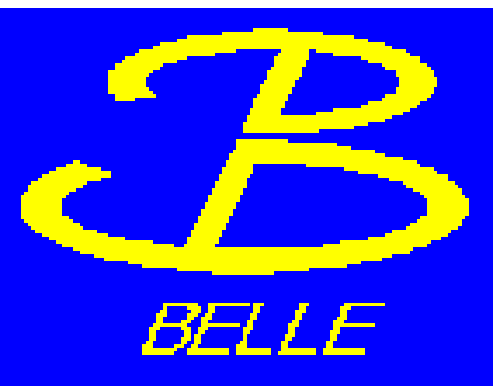}}
\preprint{\vbox{ \hbox{Belle preprint 2004-22}
                 \hbox{KEK   preprint 2004-38}
                 \hbox{hep-ex/0408051}
}}

\title{\boldmath Search for $CP$ violation in the decay $B^0\to D^{*\pm}D^\mp$}

\affiliation{Budker Institute of Nuclear Physics, Novosibirsk}
\affiliation{Chiba University, Chiba}
\affiliation{Chonnam National University, Kwangju}
\affiliation{University of Cincinnati, Cincinnati, Ohio 45221}
\affiliation{University of Frankfurt, Frankfurt}
\affiliation{University of Hawaii, Honolulu, Hawaii 96822}
\affiliation{High Energy Accelerator Research Organization (KEK), Tsukuba}
\affiliation{Hiroshima Institute of Technology, Hiroshima}
\affiliation{Institute of High Energy Physics, Chinese Academy of Sciences, Beijing}
\affiliation{Institute of High Energy Physics, Vienna}
\affiliation{Institute for Theoretical and Experimental Physics, Moscow}
\affiliation{J. Stefan Institute, Ljubljana}
\affiliation{Kanagawa University, Yokohama}
\affiliation{Korea University, Seoul}
\affiliation{Kyungpook National University, Taegu}
\affiliation{Swiss Federal Institute of Technology of Lausanne, EPFL, Lausanne}
\affiliation{University of Ljubljana, Ljubljana}
\affiliation{University of Maribor, Maribor}
\affiliation{University of Melbourne, Victoria}
\affiliation{Nagoya University, Nagoya}
\affiliation{Nara Women's University, Nara}
\affiliation{National Central University, Chung-li}
\affiliation{National United University, Miao Li}
\affiliation{Department of Physics, National Taiwan University, Taipei}
\affiliation{H. Niewodniczanski Institute of Nuclear Physics, Krakow}
\affiliation{Nihon Dental College, Niigata}
\affiliation{Niigata University, Niigata}
\affiliation{Osaka City University, Osaka}
\affiliation{Osaka University, Osaka}
\affiliation{Panjab University, Chandigarh}
\affiliation{Peking University, Beijing}
\affiliation{Princeton University, Princeton, New Jersey 08545}
\affiliation{University of Science and Technology of China, Hefei}
\affiliation{Seoul National University, Seoul}
\affiliation{Sungkyunkwan University, Suwon}
\affiliation{University of Sydney, Sydney NSW}
\affiliation{Tata Institute of Fundamental Research, Bombay}
\affiliation{Toho University, Funabashi}
\affiliation{Tohoku Gakuin University, Tagajo}
\affiliation{Tohoku University, Sendai}
\affiliation{Department of Physics, University of Tokyo, Tokyo}
\affiliation{Tokyo Institute of Technology, Tokyo}
\affiliation{Tokyo Metropolitan University, Tokyo}
\affiliation{Tokyo University of Agriculture and Technology, Tokyo}
\affiliation{University of Tsukuba, Tsukuba}
\affiliation{Virginia Polytechnic Institute and State University, Blacksburg, Virginia 24061}
\affiliation{Yonsei University, Seoul}
  \author{T.~Aushev}\affiliation{Institute for Theoretical and Experimental Physics, Moscow} 
  \author{Y.~Iwasaki}\affiliation{High Energy Accelerator Research Organization (KEK), Tsukuba} 
  \author{K.~Abe}\affiliation{High Energy Accelerator Research Organization (KEK), Tsukuba} 
  \author{K.~Abe}\affiliation{Tohoku Gakuin University, Tagajo} 
  \author{I.~Adachi}\affiliation{High Energy Accelerator Research Organization (KEK), Tsukuba} 
  \author{H.~Aihara}\affiliation{Department of Physics, University of Tokyo, Tokyo} 
  \author{M.~Akatsu}\affiliation{Nagoya University, Nagoya} 
  \author{Y.~Asano}\affiliation{University of Tsukuba, Tsukuba} 
  \author{T.~Aziz}\affiliation{Tata Institute of Fundamental Research, Bombay} 
  \author{S.~Bahinipati}\affiliation{University of Cincinnati, Cincinnati, Ohio 45221} 
% ...
\collaboration{The Belle Collaboration}

\noaffiliation

\begin{abstract}
We report a search for $CP$-violating asymmetry in $B^0\to
D^{*\pm}D^\mp$ decays. The analysis employs two methods of $B^0$
reconstruction: full and partial. In the full reconstruction method
all daughter particles of the $B^0$ are required to be detected; the
partial reconstruction technique requires a fully reconstructed $D^-$
and only a slow pion from the $\DstP\to D^0\pi_{\rm slow}^+$ decay.
From a fit to the distribution of the time interval corresponding to
the distance between two $B$ meson decay points we calculate the
$CP$-violating parameters and find the significance of nonzero $CP$
asymmetry to be $2.7$ standard deviations.
\end{abstract}

\maketitle


In the Standard Model (SM), $CP$ violation arises from the
Kobayashi-Maskawa (KM) phase~\cite{ckm} in the weak interaction
quark-mixing matrix.  Comparisons between SM expectations and
measurements in various modes are important to test the KM model.  The
\BDDst\ modes are of particular interest since large $CP$ violation
effects are expected in these decays~\cite{aleksan}.  Although the
$D^{*\pm}D^{\mp}$ final states are not $CP$ eigenstates, they can be
produced in the decays of both $B^0$ and $\bar B^0$ with comparable
amplitudes; the interference between amplitudes of the direct
transition and that via $B\bar B$ mixing results in $CP$
violation. These decays are dominated by the tree $b\to c\bar cd$
transition, thus $CP$ violation measurements are sensitive to the
angle $\phi_1$.  However, the $b\to d$ penguin diagram also
contributes to this final state and contains a different weak phase.
Therefore this contribution results in both direct $CP$ violation and
a deviation of the mixing-induced $CP$ violation parameter from
$\sin2\phi_1$.  The Cabibbo suppressed decays \BDDst\ were first
observed by Belle~\cite{belle}, and have been confirmed by
{\it BABAR}~\cite{babar}.

The probability for a $B$ meson to decay to $D^{*\pm}D^{\mp}$ at
time \Dt\ can be expressed in terms of five parameters, ${\cal A}$,
$S_\pm$ and $C_\pm$:
\begin{eqnarray} 
  {\cal P}_{D^*D}^\pm(\Dt)=(1\pm{\cal A})
  \frac{e^{-\left|\Dt\right|/\tau_{B^0}}}{8\tau_{B^0}}
  \{1+q[S_\pm\sin(\Delta m_d\Dt)-C_\pm\cos(\Delta m_d\Dt)]\}.
  \label{babarparam}
\end{eqnarray}
Here the $+(-)$ sign represents the \DstP\DM\ (\DstM\DP) final
state, and the $b$-flavor charge $q=+1(-1)$ when the tagging $B$ meson 
is a $B^0(\bar B^0)$.  The time-integrated asymmetry ${\cal A}$ 
between the rates to \DstP\DM\ and \DstM\DP\ is defined as
\begin{eqnarray}
  {\cal A}=\frac{N_{\DstP\DM}-N_{\DstM\DP}}{N_{\DstP\DM}+N_{\DstM\DP}}.
\end{eqnarray}
In the case of negligible penguin contributions~\cite{aleksan,xing},
the parameters $S_\pm$ can be related to the weak phase difference
($\sin2\phi_1$ in the SM), the strong phase difference ($\delta$) and
the ratio of tree amplitudes to the \DstP\DM\ and \DstM\DP\ final
states.  If $\delta=0$ and equal amplitudes are assumed, one expects
that ${\cal A}=0$, $C_+=C_-=0$ and $S_+=S_-=-\sin2\phi_1$.


The analysis described here is based on $140\,$fb$^{-1}$ of data,
corresponding to $152\times10^6$ $B\bar B$ pairs, collected with the
Belle detector~\cite{beldetec} at the KEKB asymmetric energy storage
rings~\cite{KEKB}.  Two reconstruction techniques, full and partial,
are used to increase the reconstruction efficiency.  The event
selection is similar to that in our previous publication~\cite{belle},
however some requirements are relaxed in order to increase the size of
the sample used to extract the $CP$ violation parameters.  The full
reconstruction method allows extraction of the signal decay with high
purity, but, due to the small branching fractions of charmed meson
decays into reconstructable final states, results in a low
efficiency. In the partial reconstruction method, a \DM\ meson is
fully reconstructed while only the slow pion (\pis) is required to be
detected from the decay $\DstP\to D^0\pis$.


Neutral $D$ mesons are reconstructed in five decay modes: $K^-\pi^+$,
$K^-\pi^+\pi^+\pi^-$, $K^-\pi^+\pi^0$, $K_S\pi^+\pi^-$ and
$K^+K^-$~\cite{cc}.  Charged $D$ mesons are reconstructed via decays
into $K^+\pi^-\pi^-$, $K_S\pi^-$ and $K^+K^-\pi^-$. The selected
combinations are fitted to a common vertex and a vertex quality
requirement is applied to reduce combinatorial background.  A
$\pm15\,$MeV$/c^2$ interval around the nominal $D$ mass is used to
select $D$ meson candidates for all modes except $D^0\to
K^-\pi^+\pi^0$, for which $\pm24\,$MeV$/c^2$ is used ($\sim3\sigma$ in
each case).  The selected charmed meson candidates are then subjected to
mass-vertex constrained fits to improve their momentum and vertex
resolution.  We refer to a $D$ candidate as having valid vertex
reconstruction if it is formed by at least two tracks with hits in the
silicon vertex detector.  To suppress feed-down from the Cabibbo
allowed decay $\bar B^0\to\DstP D^{(*)-}_s$ due to $K/\pi$
misidentification, we apply a $D_s^-$ veto for the $D^-\to
K^+\pi^-\pi^-$ and $K_S\pi^-$ channels: if a pion candidate can also
be identified as a kaon, and if, after reassignment of the kaon mass, the
invariant mass of the combination is within $\pm15\,$MeV$/c^2$ of the
nominal $D_s^-$ mass, the combination is rejected.  This requirement
suppresses the $\bar B^0\to\DstP D^{(*)-}_s$ background by a factor of
10 with signal efficiency of 98\%.  The \DstP\ candidates are formed
from $D^0\pis$ combinations with invariant masses within
$\pm2\,$MeV$/c^2$ of the nominal \DstP\ mass.

\begin{figure}[ht]
  \includegraphics[width=.7\textwidth]{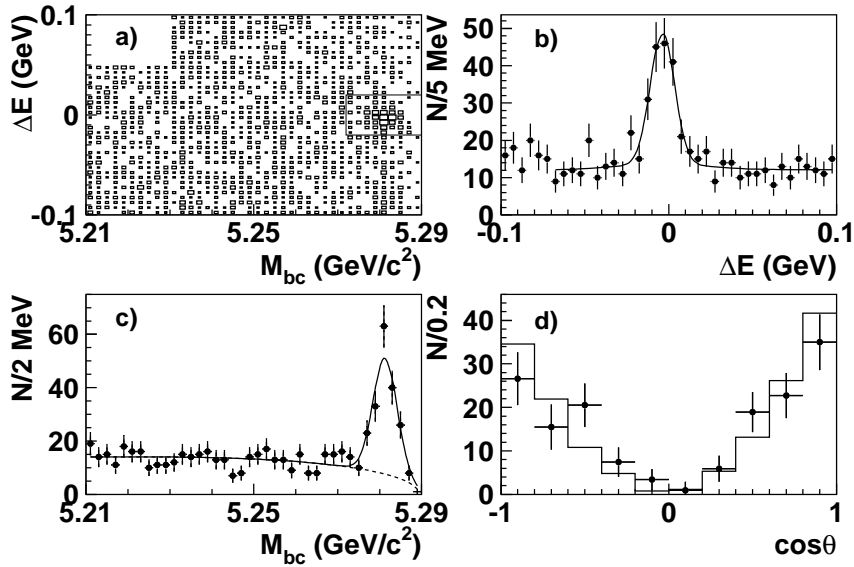}
  \caption{Kinematic distributions of \BDDst\ candidates: a) scatter
    plot of \DE\ \emph{vs.} \Mbc, b), c) \DE\ and \Mbc\ projections,
    d) \ctheta\ determined from \Mbc\ fits in the data (points with
    error bars) and in the signal MC (histogram).  The curves
    represent the fit described in the text.}
  \label{fullsig}
\end{figure}

In the full reconstruction method we define $B^0$ candidates as
combinations of oppositely charged \DstP\ and \DM\ candidates, where
at least one of the \DM\ or the $D^0$ from the \DstP\ decay has valid
vertex reconstruction.  The signal is identified using the
c.m. system energy difference $\DE=E_B^*-E_{\rm beam}$
and the beam-energy constrained mass $\Mbc=\sqrt{E_{\rm
beam}^2-P_B^{*2}}$, where $E_B^*$ ($P_B^*$) is the energy (momentum)
of the $B$ candidate in the c.m. and $E_{\rm beam}$ is the c.m. beam
energy.  $B^0$ candidates are preselected by requiring
$\left|\DE\right|<100\,$MeV and $\Mbc>5.21\,$GeV$/c^2$. In case of
multiple $B^0$ candidates in this region a single candidate per event
is selected based on the minimum sum of $\chi^2$/DOF of the fits to
intermediate charmed mesons.  The scatter plot of \DE\ \emph{vs.}
\Mbc\ and the \DE\ and \Mbc\ projections are shown in
Figs.~\ref{fullsig}.  In the \Mbc\ projection $B^0$ candidates are
selected from the $\left|\DE\right|<20\,$MeV region; the \DE\
distribution is plotted for the region $\Mbc>5.27\,$GeV$/c^2$. A fit
to the \Mbc\ distribution with a Gaussian representing the signal
contribution and an ARGUS function~\cite{argfunc} parameterizing the
background finds a signal yield of $161\pm16$ events. A fit to the
\DE\ distribution is performed using a double Gaussian to parameterize
the signal, while the background is described by a linear
function. This fit yields $149\pm18$ signal events.  The \ctheta\
distribution, where $\theta$ is a decay angle in the \DstP\ rest
frame relative to the boost direction, for the candidates from the
$\left|\DE\right|<20\,$MeV and $\Mbc>5.27\,$GeV$/c^2$ region
determined from \Mbc\ fits is shown in Fig.~\ref{fullsig} d) and is in
good agreement with the Monte Carlo (MC) expectation.


In the partial reconstruction analysis we define $B$ candidates as
combinations of \DM\ with valid vertex reconstruction and \pis.  As in
our previous publication~\cite{belle}, the angle $\alpha$ between the
\DM\ and \pis\ c.m. momenta, and the \DstP\ helicity angle $\theta$,
calculated using kinematic constraints, are used to identify the
studied decay.  We use the $\DM\to K^+\pi^-\pi^-$ decay mode only. In
addition, the \DM\ c.m. momentum is required to lie in the interval
$1.63\,$GeV$/c<P^*_{\DM}<1.97\,$GeV$/c$, and the c.m. momentum of \pis\
is required to be smaller than $0.2\,$GeV$/c$. Both momentum intervals
correspond to the kinematic limits for the studied decay. In order to
make the fully and partially reconstructed samples statistically
independent, \pis\ is rejected if, after being combined with any $D^0$
in the event, it forms a \DstP\ candidate.  The presence of a lepton
(\ltag) in the event is required to provide flavor tagging, suppress
the continuum background to a negligible level, and also reduce the
combinatorial $B\bar B$ background.  Charged tracks with c.m. momenta
in the range $1.1\,$GeV$/c<P^*_{\ltag}<2.3\,$GeV$/c$, which are identified as
muons or electrons are considered as leptons.  A large fraction of the
selected \DM\ltag\ combinations originate from the decay of the same
$B$ meson: $B\to D^-\ell^+X$.  This background is removed by a
kinematic requirement that \DM\ and \ltag\ do not originate from the
same $B$:
\begin{eqnarray}
  \frac{(E_{\rm beam}-E_{D\ltag}^*)^2-P_B^{*2}-P_{D\ltag}^{*2}}
       {2P_B^*P_{D\ltag}^*}<-1.1,
\end{eqnarray}
where $E_{D\ltag}^*$ ($P_{D\ltag}^*$) is the c.m. energy (momentum) of
the \DM\ltag\ combination.  The efficiency of this requirement for the
signal is estimated from MC simulation to be $87\%$, while the
background is suppressed by a factor greater than 2.  This
requirement also removes leptons produced from the unreconstructed
$D^0$ in the signal decay, an additional source of mistagging.  We
select partially reconstructed $B^0$ candidates by requiring
$\calpha<0$ and $\left|\ctheta\right|<1.05$. In case of multiple
candidates, the \DM\pis\ combination with the best probability of the
\DM\ vertex fit or the largest $\left|\ctheta\right|$ is selected.
The expected number of signal events in the partial reconstruction
sample is calculated from the full reconstruction signal yield relying
on the MC ratio of full and partial reconstruction efficiencies. We
estimate $N_{\rm partial}=133\pm13$ and use this number to fix the
signal fraction in later fits.

\begin{figure}[ht]
  \includegraphics[width=.7\textwidth]{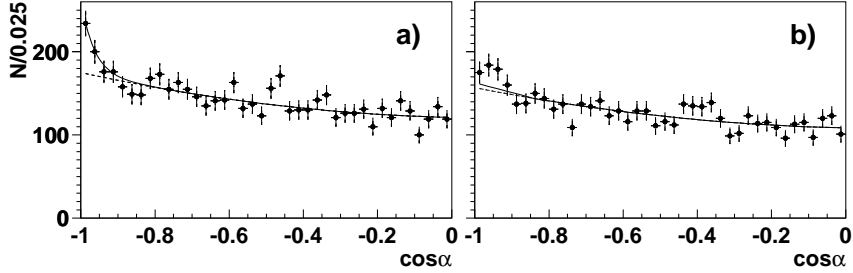}
  \caption{Distributions of $\calpha$ for: a)
    $0.50<\left|\ctheta\right|<1.05$; b)
    $\left|\ctheta\right|<0.50$. The fit functions are shown with
    solid lines; the combinatorial backgrounds are presented by dashed
    lines.}
  \label{cosdpi2all}
\end{figure}

We cross-check this result by estimating the signal fraction from the
data.  The distributions of \calpha\ for two regions of \ctheta\ are
shown in Fig.~\ref{cosdpi2all}, after imposing a tight requirement of
$\pm8\,$MeV$/c^2$ ($\sim2\sigma$) on the \DM\ mass.  The first region
$0.50<\left|\ctheta\right|<1.05$ (Fig.~\ref{cosdpi2all} a)) is signal
enriched due to the \DstP\ polarization; the second region
$\left|\ctheta\right|<0.50$ (Fig.~\ref{cosdpi2all} b)) is dominated by
background. In a simultaneous fit to the two \calpha\ distributions
the signal shapes are fixed from the MC.  The combinatorial background
is parameterized by a second order polynomial function. The
contributions from $\bar B^0\to\DstP D^{(*)-}_s$ and
$B^0\to\DstP\DstM$ are fixed from the MC simulation.  The fit yields
$137\pm39$ signal events, in good agreement with the yield expected
from the full reconstruction analysis.


In the full reconstruction method, charged tracks that are not
associated with the reconstructed \BDDst\ are used to identify the
flavor of the signal $B^0$~\cite{jpsi}.  Events are divided into six
subsamples of the parameter $r$, which is an event-by-event,
MC-determined flavor-tagging quality factor that ranges from $r=0$ for
no flavor discrimination to $r=1$ for unambiguous flavor assignment.
The wrong-tag fraction and difference between $B^0$ and $\bar B^0$
decays in each interval ($w_i$ and $\Delta w_i$, $i=1,6$) are fixed
using a data sample of self-tagged $B^0$ decay modes.  In the partial
reconstruction case, flavor tagging is provided by the high momentum
lepton required in the event; the wrong tag fraction is determined
from data as discussed below.


The proper-time difference between the reconstructed and tagged $B$
decay is calculated as $\Dt=(z_{D^*D}-z_{\rm tag})/\beta\gamma c$,
where $z_{D^*D}$ and $z_{\rm tag}$ are the $z$ coordinates of the two
$B$ decay vertices and $\beta\gamma=0.425$ is the Lorentz boost factor
at KEKB. In the full reconstruction method we determine the $B^0$
signal vertex by fitting the momentum vectors of \DM\ and/or $D^0$
candidates with well-reconstructed vertices with the constraint of the
interaction region profile.  The tagging $B$ vertex is found using
well reconstructed charged tracks not assigned to the signal $B^0$ and
excluding tracks that form a $K_S$ candidate.  The signal resolution
function parameters are obtained from the \Dt\ fit to the $B^0$
lifetime for the events from the signal region.  The tagging $B^0$
vertex resolution function is fixed from~\cite{jpsi}.  In the partial
reconstruction method, the signal and tagging $B^0$ vertices are
reconstructed using the \DM\ candidate and \ltag, respectively. In
this case, both the resolution function parameters and the wrong tag
fraction are extracted from the data using a sample of
$B^0\to\DM\ell^+\nu X$ decays tagged with a high momentum lepton.  The
$\DM\ell^+$ combinations are required to originate from the same $B$
decay based on the recoil mass against the $\DM\ell^+$ system and its
c.m. momentum. The selected $\DM\ell^+$ combinations are almost pure
$B^0\to\DM\ell^+\nu$ signal events with a small admixture of
$B^0\to\DstM(\DM\pi^0)\ell^+\nu$; the latter process is also
considered as signal. A small contribution from combinatorial
background under the \DM\ peak is estimated using \DM\ mass sidebands.
The \DM, $\ell^+$ and \ltag\ vertices ($z_{\DM}$, $z_{\ell^+}$ and
$z_{\ltag}$) are reconstructed using identical procedures to those
used in the partial reconstruction method. The resolution function is
extracted from an unbinned maximum likelihood fit to the
$\Dt_{\ell}\equiv(z_{\DM}-z_{\ell})/\beta\gamma c$ distribution.  The
wrong tag fraction is found from a fit to $\Dt_{\rm
tag}\equiv(z_{\DM}-z_{\rm tag})/\beta\gamma c$ to be $w=6.1\pm0.9\%$.
As a cross-check, the $B^0$ lifetime and mixing parameter $\Delta m_d$
are also measured from the fit to the $\Dt_{\rm tag}$ distribution to
be $1.48\pm0.04\,$ps and $0.52\pm0.02\,$ps$^{-1}$, respectively,
consistent with~\cite{pdg}.


In the full reconstruction method, the signal region is defined as
$\left|\DE\right|<50\,$MeV and $\Mbc>5.27\,$GeV$/c^2$, and contains
$360$ events with $46\%$ signal purity. In the partial reconstruction
method, the signal region is chosen as
$\left|M_{K^+\pi^-\pi^-}-M_{\DM}\right|<15\,$MeV$/c^2$, $\calpha<-0.9$
and $\left|\ctheta\right|<1.05$.  The total number of selected events
is $2174$ with $6\%$ signal purity.

We determine the $CP$ violation parameters from an unbinned maximum
likelihood fit to the \Dt\ distribution.  The signal probability
density function is given by Eq.~\ref{babarparam} with effects due to
mistagging taken into account.  The resolution function $R_{D^*D}$ is
formed by convolving four components: the detector resolutions for
$z_{D^*D}$ and $z_{\rm tag}$, the shift in the $z_{\rm tag}$ vertex
position due to secondary tracks originating from charmed particle
decays, and the smearing due to the kinematic approximation used to
convert $\Delta z$ to \Dt~\cite{jpsi}.  For each event we define the
following likelihood value
\begin{eqnarray} 
    P_{i}=\int\Big[\frac{f_{D^*D}}{f_{D^*D}+f_{\rm bg}}{\cal P}_{D^*D}(\Dt')
    R_{D^*D}(\Dt_i-\Dt') 
    \hspace*{1cm} \nonumber \\ 
    +\frac{f_{\rm bg}}{f_{D^*D}+f_{\rm bg}}
    {\cal P}_{\rm bg}(\Dt')R_{\rm bg}(\Dt_i-\Dt')\Big]d\Dt',
\end{eqnarray}
where signal ($f_{D^*D}$) and background ($f_{\rm bg}$) fractions are
calculated as functions of the following variables: \DE\ and \Mbc\
(full reconstruction); $M_{\DM}$ and \calpha\ (partial
reconstruction); \ctheta\ (both cases).  The signal distributions of
the variables used for $f_{D^*D}$ parametrization are determined from
the MC simulation.  The background parameters are obtained from the
data.

In the full reconstruction, the background \Dt\ shape is fixed using
the large \Mbc-\DE\ region excluding the signal region.  For the
partial reconstruction the background contains a combinatorial
component, for which the shape is obtained from \DM\ mass sidebands
($30\,$MeV$/c^2<\left|M_{K^+\pi^-\pi^-}-M_{\DM}\right|<60\,$MeV$/c^2$),
and a component containing a real \DM, which may come from $B$ decay.
The shape of the latter includes a mixing term, and is obtained from a
sideband ($-0.8<\calpha<0.0$).


Finally, the results for the $CP$ violation parameters ${\cal A}$,
$S_\pm$, and $C_\pm$ obtained from the individual fits to the
statistically independent full reconstruction and partial
reconstruction samples, as well as the result of the combined fit, are
summarized in Table~\ref{results}.
We calculate the combined statistical significance of $CP$ violation
to be $2.7\sigma$. The significance is defined as
$\sqrt{-2\ln(L_0/L_{\rm max})}$, where $L_{\rm max}$ is the likelihood
returned by the combined fit and $L_0$ is determined from a fit with
the parameters ${\cal A}$, $S_{\pm}$ and $C_{\pm}$ constrained to the
values corresponding to no $CP$ violation: ${\cal A}=0$, $S_+=-S_-$
and $C_+=-C_-$.
\begin{table}[tb]
  \caption{Fit results.}
  \begin{center}
    \begin{ruledtabular}
      \begin{tabular}{lccc}
	& full rec. & partial rec. & combined  \\
	\hline
	${\cal A}$ & $+0.03\pm0.09$ & $+0.16\pm0.18$ & $+0.07\pm0.08$ \\
	$S_-$      & $-1.17\pm0.48$ & $-0.65\pm0.79$ & $-0.96\pm0.43$ \\
	$C_-$      & $+0.33\pm0.29$ & $+0.11\pm0.45$ & $+0.23\pm0.25$ \\
	$S_+$      & $-0.25\pm0.52$ & $-0.92\pm0.58$ & $-0.55\pm0.39$ \\
	$C_+$      & $-0.34\pm0.27$ & $-0.39\pm0.38$ & $-0.37\pm0.22$ \\
      \end{tabular}
    \end{ruledtabular}
  \end{center}
  \label{results}
\end{table}

\begin{figure}[htb]
  \includegraphics[width=.5\textwidth]{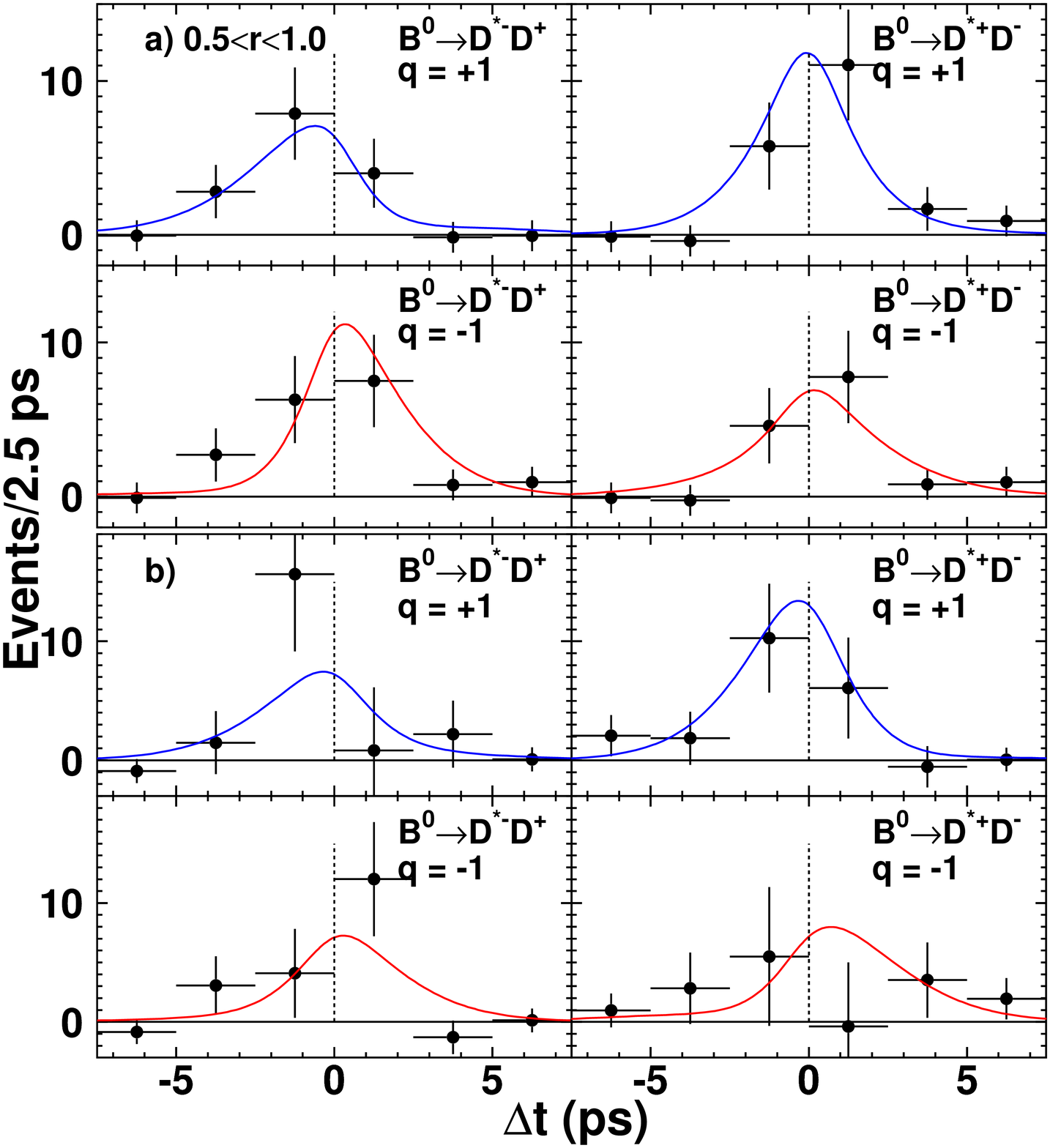}
  \caption{Background subtracted \Dt\ distributions in the a) full and
    b) partial reconstruction methods. The curves show the result of
    the fits.}
  \label{cpfit}
\end{figure}

The \Dt\ distributions for the subsamples having the best signal and
tagging quality ($f_{D^*D}>0.1$ and $r>0.5$) after background
subtraction are shown in Fig.~\ref{cpfit} a) and b) for the full and
partial reconstruction methods, respectively.


The systematic error is dominated by the uncertainties in the signal
fraction ($\pm0.07$ for $S_\pm$ and $\pm0.03$ for $C_\pm$), wrong tag
fraction ($\pm0.05$ for $S_\pm$ and $\pm0.03$ for $C_\pm$), resolution
function parameterization ($\pm0.05$ for $S_\pm$ and $\pm0.02$ for
$C_\pm$) and vertexing ($\pm0.05$ for $S_\pm$ and $\pm0.01$ for
$C_\pm$). Other contributions come from the correlated backgrounds and
signal box definition.  The result is consistent with~\cite{babar}.
We perform a number of cross-checks for our measurement. Using an
ensemble of MC pseudo-experiments, we check both the linearity of the
fitting procedure and the reliability of the statistical errors
returned by the $CP$ fit. A similar $CP$ violation study is performed
with self-tagged $\bar B^0\to\DstP D_s^-$ decay using both full and
partial reconstruction techniques.  The combined fit yields ${\cal
A}=0.00\pm0.03$, $S_-=+0.08\pm0.12$, $C_-=-1.11\pm0.07$,
$S_+=+0.00\pm0.12$, and $C_+=+1.12\pm0.07$, consistent with the
expected values ${\cal A}=0$, $S_\pm=0$ and $C_+=-C_-=1$.


In summary, we have performed a search for the $CP$-violating
asymmetry in the decay \BDDst\ using two methods of $B^0$
reconstruction.  From the combined fit to the data we have measured
${\cal A}=+0.07\pm0.08\pm0.04$, $S_-=-0.96\pm0.43\pm0.12$,
$C_-=+0.23\pm0.25\pm0.06$, $S_+=-0.55\pm0.39\pm0.12$ and
$C_+=-0.37\pm0.22\pm0.06$.  These are the most precise measurements of
these parameters to date.  The significance of nonzero $CP$ violation
in \BDDst\ is $2.7\sigma$.


We thank the KEKB group for the excellent operation of the accelerator,
the KEK Cryogenics group for the efficient operation of the solenoid,
and the KEK computer group and the NII for valuable computing and
Super-SINET network support.
We acknowledge support from MEXT and JSPS (Japan);
ARC and DEST (Australia);
NSFC (contract No.~10175071, China);
DST (India);
the BK21 program of MOEHRD and the CHEP SRC program of KOSEF (Korea);
KBN (contract No.~2P03B 01324, Poland);
MIST (Russia);
MESS (Slovenia);
NSC and MOE (Taiwan);
and DOE (USA).


\begin{thebibliography}{99}
\bibitem{cc}
  The inclusion of charge conjugate modes is implied through this Letter.
\end{thebibliography}
\end{document}